# .Black holes to white holes II:
## quasi-classical scenarios for white hole evolution

James M. Bardeen

*Physics Department, Box 1560, University of Washington*
*Seattle, Washington 98195-1560. USA*
*bardeen@uw.edu*

## Abstract

This paper considers a wider range of "quasi-classical" models for the non-singular transition from an evaporating black hole to a white hole and the evolution of the white hole than were considered in Paper I.  The quantum evolution of the geometry outside a collapsing shell is described by a smoothly evolving spherically symmetric effective metric, with a transition from trapped surfaces in the black hole to anti-trapped surfaces in the white hole at a spacelike hypersurface on which the circumferential radius has a minimum Planck-scale value.  Rather than focusing on a specific model for the effective metric and the corresponding effective stress-energy tensor, I consider the general properties of such a transition and the end of the black hole following from the assumption of a smooth geometry. Alternative scenarios for the evolution of the white hole, which avoid the prolonged emission of negative energy from the white hole of the Paper I scenario, are explored.  If such a transition, suggested by Loop Quantum Gravity, can be firmly established, the conventional holographic interpretation of the fine-grained entropy of a black hole must be abandoned.  It would also be a counter-example to the generalized second law and quantum singularity theorems.

## I. INTRODUCTION

The quantum evolution of black holes has been the subject of vigorous controversy ever since the discovery of Hawking radiation[1] 45 years ago.  The Hawking radiation is entangled with Hawking "partners."  If these are trapped inside the black hole, and if the black hole evaporates completely without release of the quantum information associated with the Hawking "partners", an initial pure state would not evolve into a pure state as determined by observers who remain outside the black hole.[2]  This black hole "information paradox" is based on the classical notion of a black hole *event* horizon sealing off any causal influence from inside the black hole on observers who remain outside the event horizon.[3]  Quantum information falling across the event horizon is swallowed up by a singularity or disappears into a "baby universe", unless it can propagate acausally.  Could it somehow all remain on or outside the event horizon even as the quantum fields propagate into the black hole[4]?  If the late Hawking radiation is entangled with early

Hawking radiation rather than with Hawking "partners" inside the horizon, purifying the state for external observers, there are predictions of a "firewall" along the event horizon[5]. For a review of some of the very extensive literature inspired by this controversy see Ref. [[6]]. This has led some to propose a drastic quantum modification of the spacetime to prevent formation of a black hole horizon (fuzzballs[7], gravastars,[8] etc.), even though quantum corrections naively should scale as $\hbar / M^2$ (units $G = c = 1$), which is incredibly small in the vicinity of the horizon of a large black hole.

The existence of a black hole *event* horizon depends on energy conditions that, while reasonable in a classical context, are known to be violated in quantum field theory. Despite this, there is a tacit assumption in most of the literature that an event horizon with an interior singularity is a general property of quantum as well as classical black holes. An elegantly written and not too technical review of attempts to understand how this might be reconciled with unitary evolution of quantum fields, as demanded, for instance, by AdS-CFT, has been posted by Almheiri, et al.[9] On the other hand, arguments have ben made based on Loop Quantum Gravity (LQG) that singularities in the black hole interior can be resolved, such that the black hole transitions into a white hole.[10,11,12,13] If this can be accomplished in a spacetime with the same causal structure as Minkowski spacetime, all quantum information trapped by the black hole eventually escapes from the white hole and reaches distant observers. If there was a pure quantum state before the black hole formed, the final result is also a pure state. It is remarkable that the possibility of such a simple resolution of the black hole information problem has been ignored by so many.

The transition from the black hole to the white hole has often been described as a "quantum tunneling" process between two quite distinct classical geometries, perhaps even before much evaporation has taken place[14]. However, Christodoulou and D'Ambrosio[15], based on a spin foam analysis in LQG, estimate the typical decay time for a spherical black hole of mass $M$ to be $M \exp\left(1820 M^2 / \hbar\right)$, which is enormously longer than the Hawking evaporation time of order $M^3 / \hbar$. If this is at all correct, it seems more reasonable to consider a "quasi-classical" model of the transition with a smoothly evolving effective metric[16]. Calculating the Einstein tensor from this metric gives an effective stress-energy tensor, which can be thought of as taking into account the average backreaction of quantum fluctuations in all of the quantum fields present. While no substitute for a full quantum gravity calculation, this can give some insight into what might dominate a quantum gravity path integral. The effective stress-energy tensor will hopefully violate classical energy conditions in a somewhat minimal way.

In a previous manuscript[17] and in Paper I[18] I described of a class of spherically symmetric models for an evaporating black hole evolving into a white hole, based on an explicit ansatz for the form of the metric, with a negative energy inflow of Hawking "partners" along ingoing radial null geodesics in the interior of the black hole. Outside the black hole horizon there is a smooth transition to outflow of positive energy Hawking radiation. The transition to the white hole is at a minimum circumferential radius $r = a$. The metric is smoothly continued into the white hole as a function of a coordinate $z$ related to the circumferential radius by $r^2 = a^2 + z^2$, and the effective stress-energy tensor has an outflow of negative energy across the white hole horizon to future null infinity. However, prolonged outflow of negative energy to large radii is problematic, since at large radii the energy density violates minimum average energy density theorems proved for quantum field theory in Minkowski spacetime.

The main purpose of this paper is to give a more general discussion of how in the quasi-classical framework the black hole can transition to the white hole. While the transition to the white hole is still assumed to be on a spacelike hypersurface at a minimum circumferential radius $r = a$ separating the trapped surfaces in the interior of the black hole from the anti-trapped 2-surfaces in the interior of the white hole, I consider alternative scenarios for evolution of the white hole, in which the white hole only emits negative energy for a limited Planck-scale range of retarded time.

The general features of a smooth quasi-classical evolution of the geometry across the transition hypersurface at $r = a$ are discussed in Part II. The geometry and effective stress-energy tensor in the neighborhood of the outer edge of the transition hypersurface, where the black hole horizon ends and the white hole horizon begins as seen by external observers, is analyzed in Part III. In Part IV I discuss alternative scenarios for the evolution of the white hole and present a detailed example. Instead of the negative energy and quantum information associated with Hawking "partners" propagating at constant retarded time across the white hole horizon over the entire lifetime of the white hole, as in the model of Paper I, most of the negative energy remains inside a white hole of fixed Planck-scale mass until it meets the rebounding matter shell, though the quantum information may continue to emerge in the form of vacuum fluctuations propagating across the white hole horizon.

The behavior of the null energy in both scenarios is the subject of Part V. In Part VI I explain in more detail than in Paper I why the standard holographic interpretation of black hole entropy, in which the log of the number of black hole quantum microstates is identified with the thermodynamic Bekenstein-Hawking entropy proportional to the horizon area, is untenable if the black hole evolves into a white hole. This has implications for the validity of the generalized second law and certain quantum singularity theorems as applied to black holes. The results of the paper are summarized Part VII, with some additional discussion.

## II. THE BASIC MODEL

The model assumes spherical symmetry of the geometry, which, while the black hole is large, is extremely close to Schwarzschild near and outside of the black hole horizon. The effective stress-energy tensor in this regime can reasonably be expected to be similar to previously calculated renormalized semi-classical stress-energy tensors for certain non-gravitational quantum fields in the Schwarzschild background[19]. The effective stress-energy tensor corrections to the geometry become substantial where the curvature becomes Planckian, at $r \sim \left(2M\hbar\right)^{2/3}$. I expect that quantum *fluctuations* in the geometry should not become large as long as $r$ is large compared with the Planck radius/mass $m_p = \sqrt{\hbar}$. In this quasi-classical regime, while the cumulative effect of the many modes of the quantum fields can be large, the contribution of any one mode to the effective geometry should be small. However, at the transition from the black hole to the white hole, as in Paper I assumed to be at a Planck-scale circumferential radius, $r^2 \sim \hbar$, quantum fluctuations in the geometry should be very large. Nevertheless, I continue the model effective metric and stress-energy tensor through the transition to a white hole. At best, this might be representative of the many trajectories contributing to a path integral for the evolution of the quantum geometry.

The key ansatz, as in Paper I and an effective geometry suggested by Ashtekar, Olmedo, and Singh[20] (AOS) on the basis of LQG, the effective metric outside the star or shell that formed the black hole has a minimum circumferential radius $r = a$ and smoothly depends on a coordinate $z$ defined by

$$r^2 = z^2 + a^2. \tag{2.1}$$

While AOS considered a fixed mass black hole, with $a$ the mass-dependent radius around where the curvature first becomes Planckian. I assume that $a^2$ is a constant of order $\hbar$, perhaps related to the fundamental LQG "area gap" parameter $\Delta$. This allows $a^2$ to be constant for an evaporating black hole. While it is desirable that the curvature not become super-Planckian, this can be accomplished by an appropriate ansatz for the rest of the effective metric.

The coordinate $z$ is defined to increase toward the future in he interior of the black hole and the interior of the white hole. In the black hole I will use advanced Eddington-Finkelstein (EF) coordinates $v, z$, which are nonsingular at the black hole apparent horizon, with a physically general line element of the form

$$ds^2 = -e^{2\psi_v} g^{zz} dv^2 - 2e^{\psi_v} dvdz + r^2 d\Omega^2. \tag{2.2}$$

The advanced time $v$ is constant along ingoing radial null geodesics, and can be initialized at past null infinity, assuming an asymptotically flat exterior. The fact that z is negative and increasing toward the future in the interior of the black hole, while $r$ is positive and decreasing toward the future, explains the difference in sign of $g_{vz}$ from the $g_{vr}$ in conventional advanced EF coordinates for the Schwarzschild geometry. Both $g^{zz}$ and $e^{\psi_v}$, as well as $r^2$, should depend on $z^2$ in the neighborhood of $z = 0$ for a smooth transition. o

In terms of these metric functions, the Misner-Sharp invariant mass function $m$ is given by, for $a^2$ a constant,

$$1 - \frac{2m}{r} \equiv \nabla_\alpha \nabla^\alpha r = \frac{z^2}{r^2} g^{zz}, \tag{2.3}$$

so

$$2m = r - \frac{z^2}{r} g^{zz}. \tag{2.4}$$

The Einstein tensor calculated from Eq. (2.2) defines the effective stress-energy tensor $T_\alpha^\beta$, with

$$8\pi T_v^v = -\frac{2}{r^2}\left(\frac{\partial m}{\partial r}\right)_v = -\frac{1}{r^2}\left[1-\left(1+\frac{a^2}{r^2}\right)g^{zz} - \frac{z^2}{r}\left(\frac{\partial g^{zz}}{\partial r}\right)_v\right], \tag{2.5}$$

$$8\pi T_v^z = \frac{2}{zr}\left(\frac{\partial m}{\partial v}\right)_z = -\frac{z}{r^2}\left(\frac{\partial g^{zz}}{\partial v}\right)_z, \tag{2.6}$$

$$8\pi e^{\psi_v}T_z^v = 2\left[\frac{a^2}{r^2} - \frac{z^2}{r}\left(\frac{\partial \psi_v}{\partial r}\right)_v\right]. \tag{2.7}$$

The $T_z^z$ component follows from the identity

$$T_z^z = T_v^v - g^{zz}e^{\psi_v}T_z^v, \tag{2.8}$$

and the conservation of the stress-energy tensor implied by the Bianchi identities gives

$$2T_\theta^\theta = \frac{1}{r}\frac{\partial}{\partial r}\left(r^2 T_z^z\right) + \frac{r^2}{z}e^{-\psi_v}\frac{\partial}{\partial v}\left(e^{\psi_v}T_z^v\right) - r\left[\frac{\partial \psi_v}{\partial r}g^{zz} + \frac{1}{2}\frac{\partial g^{zz}}{\partial r}\right]e^{\psi_v}T_z^v. \tag{2.9}$$

Constructing a particular model, for $z < 0$, means specifying $g^{zz}(v,r)$ and $\psi_v(v,r)$, which should be at least $C^1$ functions of $v$ and $r$ to ensure nonsingular curvature and a well-behaved effective stress-energy tensor. Note from Eq. (2.6) that approaching $z = 0$ the component $T_v^z$ goes to zero linearly in $z$.

The physical interpretation of the stress-energy tensor can best be assessed by projecting it onto an orthonormal tetrad of basis vectors. At the transition from the black hole interior to the white hole interior at $z = 0$, $g^{zz} < 0$, $z$ is a time coordinate, and $v$ increases going outward at constant $z$ from the collapsing star or shell that formed the black hole. A natural choice of tetrad at $z = 0$ has a 4-velocity $u^\alpha$ with $u_v = 0$, so it is orthogonal to a displacement at constant $z$. Then

$$u^v = e^{-\psi_v}/\sqrt{-g^{zz}}, \quad u^z = \sqrt{-g^{zz}}, \quad u_z = -1/\sqrt{-g^{zz}}. \tag{2.10}$$

The outward-directed radial basis vector, with $n^v > 0$, is

$$n^v = e^{-\psi_v}/\sqrt{-g^{zz}}, \quad n^z = 0, \quad n_v = e^{\psi_v}\sqrt{-g^{zz}}, \quad n_z = -1/\sqrt{-g^{zz}}. \tag{2.11}$$

In this frame, the energy density $E$, the energy flux $F$, and the radial stress $P_r$ are

$$E = -T_z^z - \left(-g^{zz}\right)^{-1}e^{-\psi_v}T_v^z = -T_z^z - F, \quad P_r = T_v^v - F. \tag{2.12}$$

Numerical calculations of the renormalized semi-classical stress-energy tensor outside the horizon of a Schwarzschild black show a flow of negative energy into the black hole starting a finite distance outside the horizon, and are inconsistent with an outflow of positive energy Hawking radiation from pair creation or tunneling very close to the horizon[21]. Inflow of negative energy along ingoing radial geodesics, like outflow of positive energy along outgoing radial null geodesics, is associated with a positive energy flux, and $T^z_v > 0$ approaching $z = 0$.

$T^z_v$ going to zero at $z = 0$ linearly in $z$ implies a change of sign of $F$ from positive in the black hole to negative in the interior of the white hole. Assuming continued flow along "ingoing" radial null geodesics, negative $F$ implies positive energy density. The contribution to the total energy of the white hole is still negative, since for "ingoing" radial null geodesics in the white hole interior, like "outgoing" null geodesics in the black hole interior, the Killing energy is negative, i.e., the energy as defined at infinity has the opposite sign from the local energy density. Another possibility is a smooth change to negative energy density *outward* flow, as in the model of Paper I.

While there are double-null coordinates that are globally non-singular, advanced EF coordinates are singular at the white hole apparent horizon, where $e^{\psi_v} \to \infty$ and changes sign. Therefore, I will switch to *retarded* EF coordinates, which are well behaved at the white hole apparent horizon. Retarded EF coordinates $u, z$ have a line element of the same form as Eq. (2.2), but with $g^{zz}(v, r) \to g^{zz}(u, r)$ and $\psi_v(v, r) \to \psi_u(u, r)$. The dependence on $r$ in the white hole may be different from that in the black hole. The retarded time is defined so that at $z = 0$, $u = -v$. The transformation from the retarded EF coordinates $x^\alpha$ with inverse metric $g^{\alpha\beta}$ to the advanced EF coordinates $\overline{x}^\mu$ with inverse metric $\overline{g}^{\mu\nu}$ is

$$\overline{g}^{\mu\nu} = \frac{\partial \overline{x}^\mu}{\partial x^\alpha} \frac{\partial \overline{x}^\nu}{\partial x^\beta} g^{\alpha\beta}, \tag{2.13}$$

and

$$\overline{g}^{zz} = g^{zz}, \quad \overline{g}_{vz} = -e^{\psi_v} = \left(\frac{\partial u}{\partial v}\right)_z g_{uz} = -\left(\frac{\partial u}{\partial v}\right)_z e^{\psi_u}. \tag{2.14}$$

At $z = 0$ $\psi_u = \psi_v$.

From the Einstein tensor in retarded coordinates,

$$8\pi T^z_u = \frac{2}{zr}\left(\frac{\partial m}{\partial u}\right)_r = -\frac{z}{r^3}\left(\frac{\partial g^{zz}}{\partial u}\right)_r. \tag{2.15}$$

The expressions for $T^u_u$ and $T^u_z$ are identical to those for $T^v_v$ and $T^v_z$ in Eqs. (2.5) and (2.7), except that the $r$ derivatives are at constant $u$ instead of at constant $v$. Near $z = 0$, $(\partial g^{zz} / \partial u)_z \cong -(\partial g^{zz} / \partial v)_z$, so $T^z_u(u, z = 0^+)$ and $T^z_v(v, z = 0^-)$ have the same sign.

The same orthonormal frame at $z = 0$ invoked earlier has basis vector components in retarded EF coordinates

$$u_u = 0, \quad u^u = e^{-\psi_u} / \sqrt{-g^{zz}}, \quad u^z = \sqrt{-g^{zz}}, \quad u_z = -1 / \sqrt{-g^{zz}}, \tag{2.16}$$

$$n^u = -e^{-\psi_u} / \sqrt{-g^{zz}}, \quad n^z = 0, \quad n_u = -e^{\psi_u}\sqrt{-g^{zz}}, \quad n_z = -1 / \sqrt{-g^{zz}}. \tag{2.17}$$

The energy density, energy flux, and radial stress in terms of the retarded coordinate components are

$$E = -T_z^{\ z} - \left(-g^{zz}\right)^{-1} e^{-\psi_u} T_u^{\ z} = -T_z^{\ z} + F, \quad P_r = T_u^{\ u} + F. \tag{2.18}$$

The sign differences in Eq. (2.18) relative to Eq. (2.12) mean that for an outward flow at constant $u$ in the white hole as described by $T_u^{\ z}$ the negative flux $F$ is associated with *negative* contributions to $E$ and $P_r$, corresponding to an outward flow of negative energy.

Note that in advanced coordinates

$$E + P_r = -T_z^{\ z} + T_v^{\ v} - 2F = -\left(-g^{zz}\right) e^{\psi_v} T_z^{\ v} - 2F, \tag{2.19}$$

while in retarded coordinates

$$E + P_r = -T_z^{\ z} + T_u^{\ u} + 2F = -\left(-g^{zz}\right) e^{\psi_u} T_z^{\ u} + 2F. \tag{2.20}$$

From Eq. (2.7), very near $z = 0$

$$e^{\psi_v} T_z^{\ v} \cong e^{\psi_u} T_z^{\ u} \cong \left(4\pi a^2\right)^{-1}, \tag{2.21}$$

implying $E + P_r < 0$. The sign of $T_z^{\ u}$ must become negative at some $z > 0$ to have positive energy density inflow inside the white hole. This requires $\left(\partial \psi_u / \partial r\right)_u > a^2 / r$ for $z > 0$, which implies that gravitational time dilation has a substantial maximum at $z = 0$, perhaps with $e^{-\psi_u} = e^{-\psi_v} \sim M / a$.

Fig. 1 shows the global causal structure for a black hole formed by the collapse of a finite thickness null shell of positive energy matter/radiation coming in from past null infinity between advanced times $v_1$ and $v_2$. The shell is assumed to bounce due to quantum backreaction and stream out between retarded times $u_2$ and $u_1$. My focus is on the geometry outside the shell. The black hole horizon is the null hypersurface at retarded time $u = 0$ for $v_0 < v < 0$. The transition from trapped surfaces in the black hole to anti-trapped surfaces in the white hole is at the minimum radius $r = a$. The white hole horizon is the null hypersurface $v = 0$ for $0 < u < u_2$. The black hole apparent horizon (i.e., trapping dynamical horizon) inside the shell is a spacelike hypersurface indicated by the lower blue line, and then while the black hole is evaporating becomes a timelike hypersurface just a Planck length or so outside the black hole horizon. The inner boundary of trapped surfaces inside the shell is indicated by the timelike upper blue line. Discussion of the white hole horizon structure is deferred for now.

The positive energy Hawking radiation, as indicated by the red arrow, flows out to future null infinity, asymptotically on outgoing radial null geodesics, over what seems an infinitesimal range of retarded times in the diagram. Of course, as measured at large radii the emission of Hawking radiation is over the very long evaporation time of the black hole, $-v_2 \sim M^3 / m_p^2$. How this outflow can be treated in advanced EF coordinates was discussed in Paper I, but it would be best handled by a switch to retarded EF coordinates somewhat outside the black hole horizon, perhaps at $r \sim 3M$. The Hawking luminosity is assumed to vanish smoothly as trapped surfaces disappear at the end of the black hole, and the metric approximately static in a small neighborhood of the $u = 0$ null hypersurface for all $v > 0$.

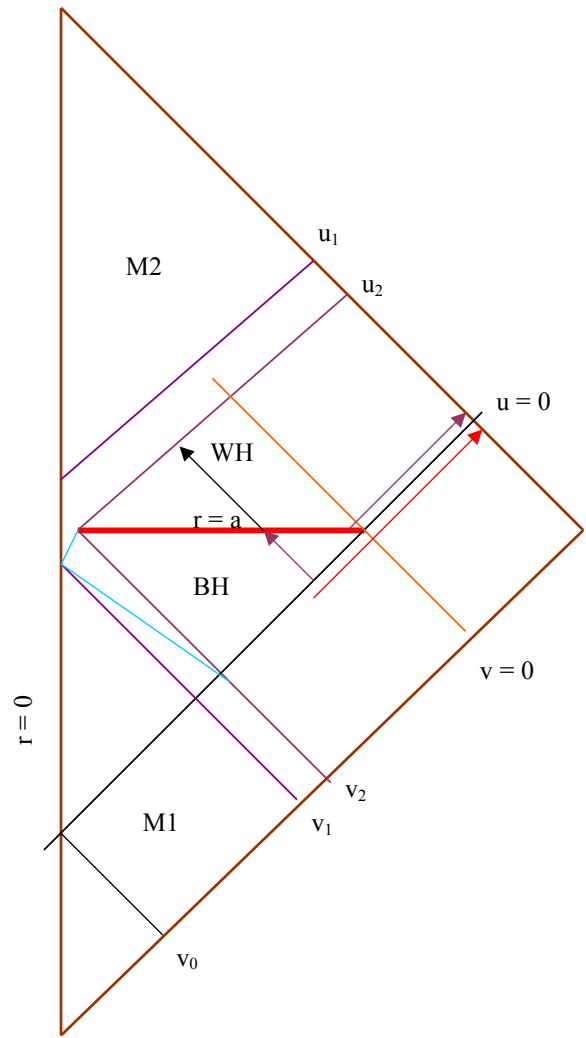

Figure 1.  A schematic Penrose diagram showing the causal strucure of a black hole to white hole spacetime.  See text for details.

The brown arrows show the negative energy associated with Hawking "partners" flowing along ingoing radial null geodesics in the interior of the black hole, and then either continuing on inward radial null geodesics in the interior of the white hole or, just after the white hole is formed, along outward radial null geodesics to future null infinity.  The flow of quantum information probably needs to be quite different, as will be explained in the discussion of the scenario presented in Part IV as an alternative to the scenario of Paper I.  The spacetime is flat in regions $M1$, $v < v_1$, and $M2$, $u > u_1$.

III.  THE END OF THE BLACK HOLE

The evaporation of the black hole plausibly continues until an advanced time at which there are no longer any trapped surfaces for $z < 0$. If evaporation were to stop sooner, the result would be a "dead" remnant black hole containing all the quantum information of the Hawking "partners." I assume a smooth endpoint of the evaporation, with the Hawking luminosity tapering off to zero as $v \to 0$, so at $z = v = 0$, $g^{zz} = 0$, $\left(\partial g^{zz} / \partial v\right)_z = 0$ and $\left(\partial g^{zz} / \partial r\right)_v > 0$. There is an outflow of positive energy Hawking radiation generated earlier crossing the $v = 0$ null hypersurface, but only for $u < 0$, from outside the black hole horizon. From Eq. (2.6), the expression for $g^{zz}$ approaching the black hole endpoint is then has the form, for $z \le 0$, $\left(z / a\right)^2 \ll 1$, and $\left(v / a\right)^2 \ll 1$,

$$g^{zz} = -A\left(v / a\right)^2 + B\left(z / a\right)^2, \tag{3.1}$$

where $A$ and $B$ are positive constants of $O(1)$. A reasonable assumption is that $\psi_v = \psi_v\left(v, 0\right) + C\left(z / a\right)^2$, with a positive constant $C$ of $O(1)$, with maximum gravitational time dilation at $z = 0$.

The effective stress-energy tensor calculated from Eqs. (2.5)-(2.9) is then

$$8\pi a^2 T_v^v = -1 - 2A\left(v / a\right)^2 + 4B\left(z / a\right)^2, \tag{3.2}$$

$$8\pi a^2 T_v^z = 2A\left(v / a\right)\left(z / a\right), \tag{3.3}$$

$$8\pi a^2 e^{\psi_v} T_z^v = 2 - 4\left(1 + C\right)\left(z / a\right)^2, \tag{3.4}$$

$$8\pi a^2 T_z^z = -1 + 2B\left(z / a\right)^2, \tag{3.5}$$

$$8\pi a^2 T_\theta^\theta = -1 + O\left(z^2 / a^2\right), \tag{3.6}$$

with a well-behaved limit as $z, v \to 0$. The black hole horizon is the null hypersurface $u = 0$, on which $\left(\partial z / \partial v\right)_u = -e^{\psi_v} g^{zz} / 2$. Approaching $v = 0$ this becomes

$$z \cong \frac{1}{6} A v^3 / a^2, \tag{3.7}$$

so the black hole horizon is inside the black hole apparent horizon at $z \cong \sqrt{A / B}\, v$, as it is for a large evaporating black hole.

Now consider the continuation to the white hole, with a switch to retarded EF coordinates. With $u = -v$ at $z = 0$ for $v < 0$, the metric in the retarded coordinates for $0 < u / a \ll 1$ and $0 < z / a \ll 1$ becomes

$$g^{zz} = -A\left(u / a\right)^2 + B\left(z / a\right)^2, \quad \psi_u = \psi_v\left(-u, 0\right) + C\left(z / a\right)^2. \tag{3.8}$$

From the metric of Eq. (3.8) the components of the effective stress-energy tensor are

$$8\pi a^2 T_u^u = -1 - 2A\left(u / a\right)^2 + 4B\left(z / a\right)^2, \tag{3.9}$$

$$8\pi a^2 T_u^z = A\left(u / a\right)\left(z / a\right), \tag{3.10}$$

$$8\pi a^2 e^{\psi_u} T_z^u = 2 - 4\left(1+C\right)\left(z/a\right)^2, \tag{3.11}$$

$$8\pi a^2 T_z^z = -1 + 2B\left(z/a\right)^2, \tag{3.12}$$

$$8\pi a^2 T_\theta^\theta = -1 + O\left(z/a\right)^2. \tag{3.13}$$

For $|u/a| \ll 1$ outside endpoint the metric and stress-energy tensor are essentially static for all $a < r < \infty$, with $T_u^z \cong 0$.

The effective stress-energy tensor is finite on the white hole apparent horizon, $u = \sqrt{B/A}\, z$ for $0 \le z/a \ll 1$. The sign of $T_u^z$ implies an initial outward flow of negative energy crossing the white hole horizon. While ingoing radial null geodesics in the interior of the black hole do have a natural continuation into the interior of the white hole, as shown by D' Ambrosio and Rovelli[22], it is not surprising that energy flow does not follow such a geodesic through the transition from the black hole to the white hole, where the geometry is changing rapidly from collapse to expansion.

The trajectories of ingoing radial null geodesics in the retarded EF coordinates are given by $\left(\partial z/\partial u\right)_v = -e^{\psi_u} g^{zz}/2$. The $v=0$ null hypersurface passing through $v = u = z = 0$ defines the white hole horizon, and for $u \ll a$ is at

$$z = \frac{1}{6} A u^3/a^2, \tag{3.14}$$

inside the white hole apparent horizon.

## IV. THE EVOLUTION OF THE WHITE HOLE

Does the initial outflow of negative energy from the white hole continue for the entire lifetime of the white hole, increasing the mass of the white hole up to the original mass of the black hole when the bouncing shell of matter that formed the black hole emerges and the white hole disappears, as assumed in Paper I? This is certainly the simplest quasi-classical model to construct. However, the negative energy density associated with the Hawking radiation only falls off as $r^{-2}$ asymptotically. Ford and Roman[23] have shown that for standard massless quantum fields in Minkowski spacetime the minimum average energy density measured by an inertial observer over a time $t_0$ is $E_{min} \sim -\hbar/t_0^4$. Taking $t_0$ to be the time tidal accelerations can be neglected at distance $r$ from a mass $M$, $E_{min} \sim -\hbar M^2/r^6$. Outflow of negative energy from the white hole that continues for very much longer than a Planck time definitely violates the Ford-Roman bound.

An alternative scenario for the evolution of the white hole is that the outflow of negative energy across the horizon lasts only for a limited retarded time, too short for the Ford-Roman bound to be violated, after which mass of the white hole becomes constant at a larger, but still Planck-scale, value until the matter shell emerges. Almost all of the negative energy associated with Hawking "partners" must then remain inside the white hole, which requires that it propagate along ingoing radial null geodesics. These increase in radius inside the white hole, and converge on the white hole apparent horizon unless they intersect the matter shell first.

A feature of a white hole apparent horizon is exponentially increasing blueshifts due to its negative surface gravity, but this is not a true instability, at least classically, since the exponentially increasing energy density is precisely compensated by decreasing volume from the convergence of geodesics toward the apparent horizon. Both are artifacts of viewing the geodesics from an accelerating frame rather than a local inertial frame. The same thing happens in a uniformly accelerating frame near the past horizon of a Rindler wedge in Minkowski spacetime. A left-going null geodesic at a small constant distance from the past Rindler horizon in the inertial frame is described in the accelerating frame as converging toward the horizon with exponentially increasing blueshift. Of course, there are only local inertial frames in the white hole geometry, as opposed to the global inertial frames of Minkowski spacetime.

The *apparent* horizon of the white hole is by definition where ingoing radial null geodesics are marginally anti-trapped, i.e., where $\left(\partial z / \partial u\right)_v = 0$. The null condition for an ingoing radial null geodesic in the retarded coordinates implies $g^{zz} = 0$ at the white hole apparent horizon.

Anti-trapping inside the white hole means both $\left(\partial z / \partial v\right)_u$ and $\left(\partial z / \partial u\right)_v$ are both greater than zero inside the white hole apparent horizon, while $\left(\partial z / \partial u\right)_v$ is zero on the apparent horizon and should be negative outside. From Eq. (2.14), with finite $\psi_u$, the identity

$$\left(\frac{\partial u}{\partial v}\right)_z = -\frac{\left(\partial z / \partial v\right)_u}{\left(\partial z / \partial u\right)_v} \qquad (4.1)$$

then implies that the metric in advanced EF coordinates is singular at the apparent horizon, with $e^{\psi_v}$ going from $+\infty$ just inside to $-\infty$ just outside, If the advanced EF metric is *not* singular at the white hole apparent horizon, then instead of $z$ (and $r$) continuing to increase beyond the white hole apparent horizon along an outgoing radial null geodesic, $z$ starts decreasing and asymptotically goes to zero as $v \to \infty$. This behavior describes a traversable wormhole with an $r = a$ throat at $v = \infty$, and is inconsistent with asymptotic flatness outside the white hole (see Simpson, et al[24]).

A simple pasting together of regions with Vaidya geometries, in which $g^{zz} = 1 - 2M/r$, with $M = M(v)$ or $M = M(u)$, as proposed in a recent paper by Martin-Dussaud and Rovelli[25], is not satisfactory as a BH-to-WH model. It doesn't deal with how the geometry transitions between the black hole and the white hole. Furthermore, the part of the semi-classical stress-energy tensor in the interior of the black hole associated with the negative energy influx of Hawking partners only increases as $r^{-2}$, while the semi-classical trace anomaly, larger just outside the black hole horizon, increases as $r^{-6}$ and quickly becomes an enormously larger contributor to the effective stress-energy tensor.

The evolution of the white hole in Paper I is very simple. Its mass increases steadily as a function of retarded time due to negative energy flowing out across the white hole horizon, in a time-reverse of the evolution of the black hole. The white hole disappears when the rebounding matter shell emerges, leaving nothing behind. There is no significant flow of energy *along* the white hole hole horizon.

To avoid trouble with the Ford-Roman energy density bound at large radii the geometry must be static outside the white hole apparent horizon for almost all of its lifetime, after the initial outflow of negative energy radiation noted in Part III. If the radius of the horizon is at least several times $a$, this static geometry should be close to Schwarzschild.

The metric ansatz adopted in Paper I has

$$g^{zz} = 1 - \left[ 2Mr^2 + \alpha ra^2 \right] / \left[ r^3 + \beta ra^2 + \gamma (2M)a^2 \right] \tag{4.2}$$

and

$$e^{-\psi_u} = e^{-\psi_v} = 1 + \delta a^2 / (2Mr) + \varepsilon a^2 / r^2 + \phi (2Ma^2) / r^3, \tag{4.3}$$

with $\alpha, \beta, \gamma$ and $\delta, \varepsilon, \phi$ constant parameters of order unity. With $\alpha = \beta + \gamma$, the $g^{zz} = 0$ condition at an apparent horizon becomes just $r = 2M$. Alternative evolution models for the white hole can be constructed by taking, instead of $M = M(u)$, $M = M(u, r)$ and modifying Eq. (4.3) appropriately. With $M = M(v)$ in the black hole, continuity requires $M(u, r = a) = M(v = -u)$ and $\psi_u = \psi_v$ at $r = a$. The beginning of the white hole apparent horizon is at $u = v = 0$, where $r = a$ and with $\alpha = \beta + \gamma$

$$2M / a = 2M_0 / a = 1. \tag{4.4}$$

A physically acceptable model must have $0 < \gamma < 1$ and $\beta > -1$.

An expression for $M(v)$ consistent with a smooth end to black hole evaporation and the semi-classical expression for the black hole luminosity at $-v = u \gg a$ implies that at $z = 0$

$$M(u) = M_0 \left[ 1 + c(u / a)^2 \right]^{1/6}, \tag{4.5}$$

with $c > 0$ of order 1 or less. Since $dM / du > 0$, there is an initial outward flow of negative energy across the white hole horizon, and once $z$ is several times $a$, $4\pi r^2 T_u^z \cong dM / du$.

A simple very ad hoc model consistent with limited emission of negative energy from the white hole is to use Eq. (4.5) for $u \le u_t$, where $u_t$ is not extremely large compared with $a$, after which $M = M(u, z)$, interpolating between $M(u)$ at $z = 0$ to a constant $M_t$ for $z \ge z_t$ in the limit $u \gg u_t$. $z_t$ is defined by $r(z_t) = 2M_t$, so $g^{zz}(u, z_t) = 0$ when $M = M_t$. Define an interpolating function

$$f(z / z_t) = 5(z / z_t)^8 - 4(z / z_t)^{10} \tag{4.6}$$

for $0 \le z \le z_t$ with $f = 1$ for $z > z_t$. Let

$$M(u,z) = M_t + \left[1 - \tanh^2(u/u_t - 1)f\right]\left[M(u) - M_t\right] \tag{4.7}$$

The stress-energy tensor obtained by patching together Eq. (4.5) and Eq. (4.7) is continuous at $z = 0$, $z = z_t$, and $u = u_t$. In the limit $u \gg u_t$ the metric becomes static on and outside the white hole apparent horizon, with no energy escaping to $\Im^+$.

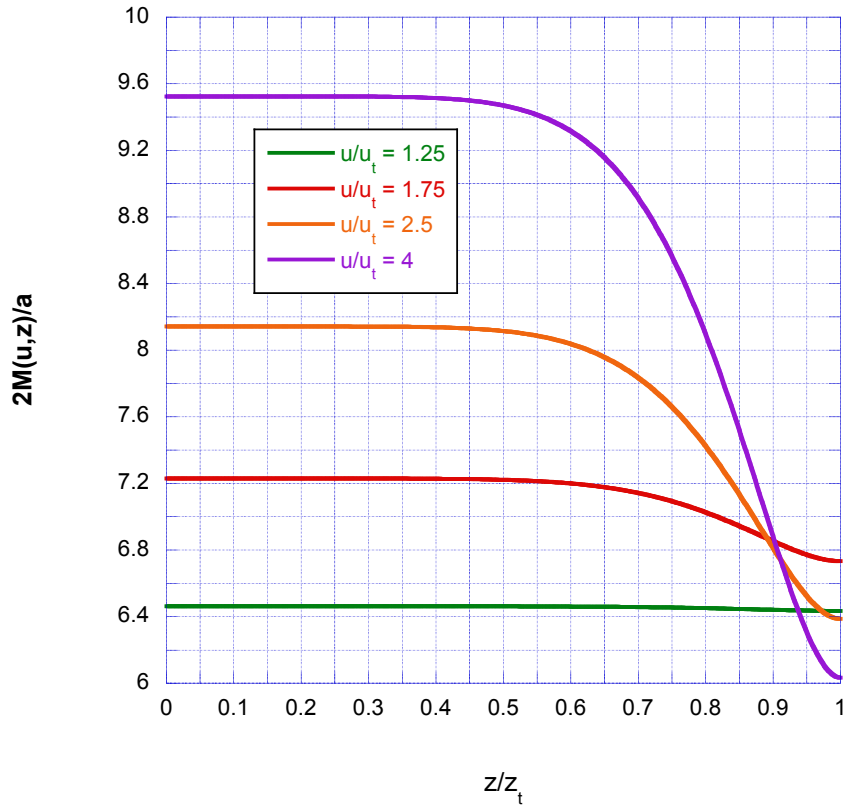

Figure 2. The model evolution of the mass parameter $M$ inside the white hole horizon for $u > u_t$, with $u_t / a = 216$ and $2M_t / a \cong 6$. In the model, $M$ is constant at a given $u$ for all $z > z_t$, and the apparent horizon is at $r = 2M$.

After $u = u_t$, the mass of the white hole at first increases, while it continues to emit negative energy, but then with some of outflow of positive energy relaxes back toward a constant $M = M_t$ (see Fig. 2). The value of the mass parameter in the interior of the white hole continues to increase, since it corresponds to the mass of the black hole at earlier and earlier advanced times $v = -u$.

Essentially *all* of the ingoing radial null geodesics from the transition hypersurface converge on the white hole apparent horizon. The equation for the trajectory is

$$\left(\frac{\partial z}{\partial u}\right)_v = -\frac{1}{2}e^{\psi_0}g^{zz}. \tag{4.8}$$

The geodesic doesn't intersect the outgoing the matter shell until $u \sim M_{sh}^3 / a^2$, where $M_{sh} \gg a$ is the mass of the shell as it formed the black hole. For any reasonable $e^{\psi_u}$, the geodesic converges toward the apparent horizon with an enormously shorter e-folding time $\sim 4 M_t$. Some of the ingoing geodesics may cross the apparent horizon while its radius is deceasing, but then they just converge on the apparent horizon from the outside. Almost all of the energy associated with Hawking "partners" ends up in a very thin shell right at the apparent horizon. The choice for $f(z / z_t)$ in Eq. (4.6) and Fig. 2 is a rather crude attempt to suggest this. The conformal rescaling in the Penrose diagram of Fig. 1 completely obscures the convergence.

It is instructive to consider the transformation of the stress-energy tensor from retarded to advanced EF coordinates. The transformation of the retarded EF metric $g_{\alpha\beta}$ to the advanced EF metric $\tilde{g}_{\alpha\beta}$ gives $\tilde{g}^{zz} = g^{zz}$ and $e^{\psi_v} = -\left(\partial u / \partial v\right)_z e^{\psi_u}$. The advanced coordinate stress-energy tensor $\tilde{T}_\alpha^\beta$ has

$$\tilde{T}_v^v = T_u^u + 2\left(g^{zz}\right)^{-1} e^{-\psi_u} T_u^z, \quad \tilde{T}_z^z = T_z^z - 2\left(g^{zz}\right)^{-1} e^{-\psi_u} T_u^z, \tag{4.9}$$

$$e^{-\psi_v} \tilde{T}_v^z = -e^{-\psi_u} T_u^z, \quad e^{\psi_v} \tilde{T}_z^v = e^{\psi_u} T_z^u + 4\left(g^{zz}\right)^{-2} e^{-\psi_u} T_u^z. \tag{4.10}$$

Inside the $g^{zz} = 0$ apparent horizon of the white hole, $\left(\partial u / \partial v\right)_z < 0$ and goes to minus infinity at the apparent horizon. Approaching the apparent horizon from outside, $\left(\partial u / \partial v\right)_z > 0$ and goes to plus infinity. Clearly, with $g^{zz}$ vanishing linearly in $z$, $e^{\psi_v} \tilde{T}_z^v$ is singular unless $e^{-\psi_u} T_u^z$ vanishes at least quadratically right at the apparent horizon. This consequence of the advanced EF coordinate singularity cannot be removed by just a uniform rescaling of $v$, under which $e^{-\psi_v} T_v^z$ is invariant.

For the stress-energy tensor in the interior of the white hole to be close to that derived from an ingoing Vaidya metric, as assumed in Ref. [25], requires $\tilde{T}_v^v \cong \tilde{T}_z^z \cong e^{\psi_v} \tilde{T}_z^v \cong 0$. This is not at all the case for my model, and doesn't take into account the large trace anomaly in the semi-classical region of the black hole interior.

The change of my retarded time coordinate $u$ over the lifetime of the white hole is by definition equal to the change of the advanced time $v$ over the lifetime of the black hole. However, the lifetime of the white hole is physically measured by the proper time of an observer at a fixed large radius. A reasonable estimate is that

$$\psi_u\left(z = 0, u\right) = \psi_v\left(0, v = -u\right), \quad e^{\psi_v(0,v)} \sim \frac{a}{2 M(v)} \sim \left(\frac{a}{u}\right)^{1/3} \tag{4.11}$$

for $u \gg a$ and $a^2$ not too large compared with $\hbar$. The physical lifetime of the white hole $\Delta \bar{u}$ is then

$$\Delta \bar{u} \sim \int_a^{\Delta v} \left(\frac{a}{u}\right)^{1/3} du \sim a^{1/3} \left(\Delta v\right)^{2/3} \sim \frac{M_{sh}^2}{a}, \tag{4.12}$$

much shorter than the lifetime $\Delta v \sim M_{sh}^3 / \hbar$ of a large black hole.

What happens when the shell at the white hole apparent horizon intersects the rebounding matter shell? Some very crude guidance may be given by the result of Dray and 't Hooft[26] for the collision of two null shells separating Schwarzschild geometries. The Schwarzschild mass is $M_t$ outside the white hole horizon, roughly $M_{sh}$ between the matter shell and the white hole horizon, and zero inside the ingoing shell after the collision. Dray and 't Hooft show that conservation laws determine the mass $M_2$ between the ingoing and outgoing shells after the collision at circumferential radius $r_0$, with (see Fig. 3)

$$\left(2 M_2 - r_0\right)\left(2 M_{sh} - r_0\right) = \left(0 - r_0\right)\left(2 M_t - r_0\right). \qquad (4.13)$$

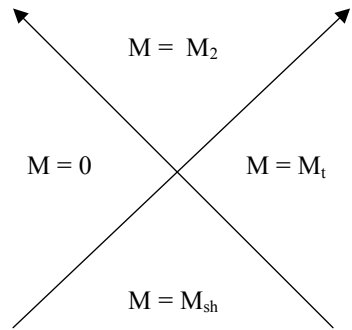

Figure 3. The intersection of the rebounding matter-radiation shell and the shell at the white hole horizon.

Since $M_1$ is a Planck scale mass, $r_0$ is very slightly less than $2 M_1$, and for an initially large black hole $M_{sh} \gg M_t$ Eq. (4.13) says that $2 M_2$ is doubly infinitesimally less than $r_0$. The mass of the outgoing shell after the collision, $M_t - M_2$, is a tiny fraction of a Planck mass, and the ingoing shell after the collision immediately forms a black hole of mass $M_2$. This classical description of the end of the white hole is highly suspect, in view of the extremely sub-Planck-scale masses and differences in circumferential radius. A full treatment in quantum gravity is really required. The analysis in Ref. [25] of the interaction of the inflowing negative energy with the outgoing matter shell does not recognize the formation of a residual black hole. The anti-trapping of two-surfaces inside the white hole ends in the expanding matter shell.

The Dray - 't Hooft result does raise serious questions about the fate of the quantum information trapped by the black hole. It is very hard to see how any but a tiny fraction of the quantum information deposited in the white hole can be contained in what little is left of the matter shell or in a residual black hole with a Planck-scale mass. The residual black hole would evaporate quickly, with the resulting white hole able to follow the scenario of Paper I without violating the Ford-Roman bound. However, the residual black hole does not have the large internal volume of the original black hole, and could not plausibly have a von

Neumann entropy vastly exceeding its small Bekenstein-Hawking entropy. As I see it, the only way this scenario can make sense is if essentially all the von-Neumann entropy of the original black hole at the end of its evaporation can gradually escape across the white hole horizon over the entire lifetime of the white hole. Zero energy vacuum fluctuations are constantly propagating on outgoing radial null geodesics across the white hole horizon, and in the presence of gravity there is not a unique vacuum state. These fluctuations, perturbed by interaction with the energy propagating along the horizon, can plausibly drain quantum information from the white hole without draining energy.

## V. NULL ENERGY CONDITIONS

The averaged (ANEC)[27] and quantum (QNEC)[28] null energy conditions were shown in Paper I to be satisfied for that paper's scenario. Here I just consider how these results might be modified in the current scenario for the evolution of the white hole. The ANEC is

$$\int_{-\infty}^{\infty} T_{\alpha\beta} k^{\alpha} k^{\beta} \, d\lambda \geq 0, \tag{5.1}$$

with $\lambda$ an affine parameter. It has been proven in some generality to hold for *achronal* null geodesics, that is, null geodesics no two points of which can be connected by a timelike curve. In a spherically symmetric geometry radial null geodesics are guaranteed to be achronal.

For an ingoing radial null geodesic the first part of the integral up through the transition to the white hole is negative, due to a dominant negative contribution from the immediate neighborhood of $z = 0$. In the retarded EF coordinates of the white hole,

$$T_{\alpha\beta} k^{\alpha} k^{\beta} d\lambda = \left[ -T_u^z - \frac{1}{4} \left( e^{\psi_u} g^{zz} \right)^2 T_z^u \right] e^{\psi_u} k^u du = \left[ -T_u^z - \left( \frac{dz}{du} \right)^2 T_z^u \right] e^{\psi_u} k^u du. \tag{5.2}$$

In the new scenario there is even more rapid exponential growth of $e^{\psi_u} k^u$ in the white hole than the original scenario, since from the geodesic equation

$$d\left( e^{\psi_u} k^u \right) / du = \left( e^{\psi_u} g^{zz} \right)_{,z} \left( e^{\psi_u} k^u \right) / 2 \tag{5.3}$$

as the geodesic approaches the white hole apparent horizon at $r = r_t \simeq 2 M_t$ the growth rate goes roughly as the constant $\left( 4 M_t \right)^{-1}$ and rather than decreasing with a steadily increasing $M(u)$, while $dz / du$ goes exponentially to zero at the corresponding rates. The key difference in the new scenario is that instead of $-T_u^z$ falling off relative slowly as a power law in $u$, it falls off as $\left( z_t - z \right)^2$, with the result that the contribution to the ANEC *falls* exponentially along the white hole horizon. The interaction with the expanding matter shell is complicated, but it seems unlikely that it can salvage the ANEC, as it did in the old scenario. The question then is whether satisfying the ANEC is really necessary for geodesics that pass through a highly quantum region of spacetime.

The QNEC is was shown to be satisfied for the Paper I scenario and is valid by a wider margin for the new scenario, in which $T_{\alpha\beta}k^\alpha k^\beta \cong 0$ on the white hole horizon instead of negative as in the old scenario.

## VI. ENTROPY

Entropy has played a central role in discussions of the evolution of black holes since before the discovery of Hawking radiation. The discovery of Hawking radiation showed that a black hole radiates like a black body and has a thermodynamic (Bekenstein-Hawking) entropy proportional to the area of the black hole horizon, $S_{BH} = A/(4\hbar) = 4\pi M^2/m_p^2$. For a sub-system of a larger quantum system, its thermodynamic entropy is usually identified with its number of quantum degrees of freedom, i.e., the logarithm of the number of its quantum states, and corresponds to its maximum possible von Neumann entropy as part of a larger system. The assumption, called the "central dogma" in Ref. [9], that this is true for black holes is what leads to the black hole information problem.

Assuming a pure quantum state before formation of the black hole, and with Hawking radiation in a completely mixed state, as predicted semi-classically, the von Neumann entropy $S_{vN}$ of the black hole increases steadily as the black hole evaporates, since the Hawking radiation is entangled with Hawking "partners" falling into the black hole, while the horizon area and $S_{BH}$ decrease. At the Page time[29], the two entropies are equal, which is when the black hole has lost about 1/2 of its initial mass. Assuming $S_{vN}$ cannot exceed $S_{BH}$, $S_{vN}$ must somehow decrease after the Page time if the evaporation of the black hole continues, and vanish as the black hole evaporates down to the Planck scale and disappears. This the famous Page curve. Somehow, the quantum information apparently carried into the black hole by Hawking partners must reappear outside the black hole horizon, even though this would seem to require acausal propagation of quantum information.

Many ideas for how a pure state can be restored, as apparently required, for instance, by the AdS/CFT correspondence, have been proposed over the years. Could *complete* quantum information be stored on the black hole horizon, at least for external observers? Could a black hole horizon never form, instead becoming a fuzzball[7] or gravastar[8] that has no event horizon and no trapped surfaces? Does entanglement between late Hawking modes and early Hawking modes apparently required to restore unitarity imply a singular "firewall" at the horizon? All of these alternatives contradict semi-classical expectations for large black holes. Why should an black hole horizon behave differently than an ordinary null hypersurface in Minkowski spacetime? After all, there is a coordinate transformation that puts the Schwarzschild metric into Rindler form for $r/(2M) - 1 \ll 1$. The latest fashionable idea, reviewed at length in Ref. [9] and based on the central dogma, is very crudely that later in the evaporation process the Hawking partners must correspond to different degrees of freedom than the earlier Hawking partners, and given the finite Hilbert space the Hawking modes with which they are entangled must then be entangled with early Hawking modes, leading to a pure external state when the black hole disappears.  This is justified by a very complicated quantum path integral involving "replica wormholes".

In any BH to WH scenario of the type considered in this paper and Paper I the central dogma is clearly false.  The entanglement entropy of the black hole is not at all constrained by the Bekenstein-Hawking entropy at late stages of black hole evaporation.  Quantum information propagating into the black hole early on crosses the $z = 0$ hypersurface into the white hole at an enormous spacelike separation from the late stage black hole horizon, as is clear from Fig. 1.  This is true regardless of whether the quantum information propagates in the interior of the black hole along ingoing radial null geodesics, as I assume, or along "outgoing" radial null geodesics.  An "outgoing" null geodesic a non-infinitesimal distance inside the horizon reaches the deep interior of the black hole in an interval of advanced time of order $M$.  Also, the matter shell that formed the black hole and any matter subsequently accreting into the black hole may have been entangled with quanta that remain outside the black hole.  The implication is that $S_{\text{BH}}$ as a thermodynamic entropy represents the maximum number of quantum degrees of freedom on the black hole horizon and able to interact with the outside world at a given advanced time, but it has nothing to do with the quantum degrees of freedom in the interior of a black hole, which are not in any kind of thermal contact with the horizon.  Rather similar arguments have been made by Garfinkle[30] and by Rovelli[31], among others.

While this behavior of the entanglement entropy of evaporating black holes violates the holographic principle as it relates to black hole horizons, there is no conflict with the Bousso covariant entropy bound[32] that Bousso used to motivate the holographic principle[33].  The covariant entropy bound states that the entropy $S$ crossing a converging null sheet orthogonal to a 2-surface of area $A$ satisfies $S \leq A / \left( 4\hbar \right)$.  For a 2-surface on a black hole horizon the null sheet in the interior of the black hole in my scenarios contains relatively little entropy, and little to none of the entropy of the earlier Hawking "partners."  The continuation of the null sheet into the white hole is expanding, not converging, so the Bousso bound does not apply there.  The conventional Page curve for the entanglement entropy has a maximum when $S_{\text{vN}} = S_{\text{BH}}$.  A BH to WH model has a similar Page curve, but with a maximum entanglement entropy at the end of black hole evaporation, when the black hole has a Planck-scale area.

The density of quantum degrees of freedom in the interior of the black hole plausibly never should exceed one per Planck volume on a maximal hypersurface.  The $z = 0$ hypersurface, which is approximately maximal except very near the end of the black hole, has a volume $dV$ for an interval of advanced time $dv$ equal to $4\pi a^2 e^{\psi_v} dv$.  A rough estimate of the number of degrees of freedom in this volume is somewhat less than

$$dN \sim L_{\text{H}} dv / T_{\text{H}} \sim \left( \hbar / M^2 \right) dv / \left( \hbar / M \right) \sim dv / M, \qquad (5.4)$$

where $M$ is the mass of the black hole at advanced time $v$ as evaluated at the black hole horizon.  The number of degrees of freedom per Planck volume is

$$\hbar^{3/2} dN / dV \le \hbar^{3/2} / \left( a^2 M e^{\psi_v} \right) \sim \left( \hbar / a \right)^{5/2},$$  (5.5)

assuming time dilation no larger than $e^{-\psi_v} \sim \hbar M / a^3$. The entropy density bound is satisfied if $a^2 / \hbar \ge O(1)$. The enormous length of the cylindrical interior of the black hole compensates even for a Planck-scale minimum circumferential radius.

Many discussions of the quantum evolution of black holes have been based on the "generalized entropy" evaluated on a closed two-surface separating "interior" and "exterior" regions of a spacelike hypersurface. The (microscopic) generalized entropy is

$$S_{\text{gen}} = \frac{A}{4G\hbar} + S_{\text{out}},$$  (5.6)

Here $A$ is the area of the two-surface, $G$ is the gravitational constant, not set equal to one since it is subject to quantum renormalization, and $S_{\text{out}}$ is the von Neumann entropy of the exterior region. The combination is thought to be finite without the need for renormalization. The *generalized second law* (GSL), originally proposed by Bekenstein[34] for black hole horizons, is that $S_{\text{gen}}$ is non-decreasing along any future causal horizon (FCH). A rather general proof of the GSL has been claimed by Wall.[35] A FCH is a null hypersurface that is the boundary of the past of a future-infinite timelike observer. While a black hole horizon, even if it is not an event horizon, at least forms part of a FCH, a white hole horizon does not. The black hole horizon in my BH to WH scenarios is consistent with the GSL.

The "outgoing" null hypersurface orthogonal to a trapped 2-surface in the interior of a black hole that evolves smoothly into a white hole is a FCH, since the null hypersurface emerges from the white hole and forms the boundary of the causal past of accelerating observers that reach $\mathfrak{I}^+$ at the same retarded time. The decrease of the area of the FCH from a macroscopic value down to the Planck scale as it approaches the transition to the white hole dominates what can only be a modest increase in $S_{\text{out}}$. Therefore, $S_{\text{gen}}$ decreases substantially while the hypersurface is in the black hole, violating the GSL and only starts increasing at the transition to the white hole.

The BH to WH scenarios are counter-examples to the Wall Quantum Singularity Theorem[36], which assumes the GSL, and to the Quantum Focusing Conjecture of Bousso, et al[37], which implies the GSL.

## VII. SUMMARY AND DISCUSSION

I have demonstrated that a smooth quasi-classical transition from an evaporating black hole to a white hole is consistent with a spacetime with the global causal structure of Minkowski spacetime, and does not require any large quantum effects outside regions of Planckian spacetime curvature. Whether this is actually consistent with quantum gravity remains an open question, but the resolution of the $r = 0$ Schwarzschild singularity and the transition from the black hole to the white hole does have support from LQG, as argued in Ref. [20]. Of course, another important unanswered question is whether any of this makes sense in the absence of strict spherical symmetry.

I have considered two quite different scenarios for the evolution of white hole in Paper I and this paper. In Paper I, I assumed the initial outflow of negative energy from the white hole associated with Hawking "partners" just after the transition from the black hole continues for the entire lifetime of the white hole. It is by far the most straightforward way to dispose of the large amount of negative energy that accumulates inside the black hole as the black hole evaporates. However, at large radii, where the geometry is close to Minkowski, the negative energy density associated with the flux of negative energy violates the Ford-Roman[22] minimum average energy density theorems. While I presented an argument in Paper I as to why the Ford-Roman bound may not be applicable to BH-to-WH scenarios, this is a serious unresolved issue.

In this paper I assume that almost all of the negative energy associated with Hawking "partners" remains inside the white hole until it interacts with the rebounding matter that formed the black hole. This requires that the metric function $g^{zz}$ in retarded EF coordinates does not depend on retarded time at and outside the white hole apparent horizon over almost all of its lifetime, even though it must do so just after the white hole is formed. The initial burst of negative energy outflow is in accord with the result of Bianchi and Smerlak[38] that, at least in 2D, unitarity cannot be preserved in the evolution of a black hole without some negative energy reaching future null infinity. The bulk of the negative energy in this scenario converges on and propagates along the white hole horizon, suggesting an approximation to the interaction with the expanding matter shell as the crossing of two infinitesimally thin null shells separating regions with Schwarzschild geometry. However, it does seem necessary that entanglement entropy continuously flows out of the white hole, in the form of vacuum fluctuations. The Dray - 't Hooft result relating the masses in the regions between crossing null shells suggests that the negative energy of the Hawking partners is absorbed by the rebounding matter shell, reducing its mass down almost to zero. The residual ingoing null shell has a Planck-scale positive mass very close to the mass of the white hole, and forms a Planck-scale black hole. This residual black hole will evaporate quickly and again evolve into a white hole, but one that can emit all its negative energy without any conflict with Ford-Roman limit on negative energy density at large radii. However, this all seems rather contrived, and the violation of the ANEC may be an issue.

BH to WH scenarios, provided they can be given a solid foundation in quantum gravity, give a simple resolution of the black hole information problem and are counter-examples to propositions that form the basis of much of the work on quantum properties of black holes, such as the generalized second law and theorems predicting singularities in the interior of quantum black holes, as discussed in Part VI. These theorems claim to assume nothing about the unknown behavior of quantum gravity in regions of very high curvature, but at least this paper's scenario is broadly consistent with semi-classical quantum field theory everywhere the spacetime curvature is less than the Planck scale. Interesting recent papers[39] on black hole evaporation in the context of AdS/CFT, while they assume the black holes start releasing quantum information at the conventional Page time, could perhaps be adapted to a BH to WH scenario with a maximum von Neumann entropy at the end of black hole evaporation. Attempts to understand the release of quantum information from macroscopic black holes in terms of quantum path integrals seem to require very exotic intermediate configurations of multiple wormholes[40], but these ignore the possibility of a much simpler path integral in a black hole to white hole scenario.

ACKNOLEDGEMENTS

The research that led to this paper was inspired by discussions with Hal Haggard while we were both visiting the Perimeter Institute in May, 2018. The Perimeter Institute is supported by the Government of Canada through the Department of Innovation, Science, and Economic Development and by the Province of Ontario through the Ministry of Research and Innovation. I also thank Amos Ori, Tommaso De Lorenzo, Pierre Martin-Dussaud, Bob Wald, and Andreas Karch for stimulating exchanges that have helped in refining my arguments.